\documentclass{IEEETran}

\usepackage{booktabs} 
\usepackage{amsmath}
\usepackage{graphicx}
\usepackage{epstopdf}
\usepackage{graphics}
\usepackage{array}
\usepackage{float}
\usepackage{multirow}

\makeatletter

\providecommand{\tabularnewline}{\\}
\floatstyle{ruled}
\newfloat{algorithm}{tbp}{loa}
\providecommand{\algorithmname}{Algorithm}
\floatname{algorithm}{\protect\algorithmname}

\makeatother

\begin{document}

\title{Towards Lightweight Error Detection Schemes for Implementations of MixColumns in Lightweight Cryptography}

\author{Anita Aghaie, Mehran Mozaffari Kermani, Reza Azarderakhsh}
\IEEEspecialpapernotice{\thanks{Anita Aghaie is with Embedded Security Group, Horst Gortz Institute for IT-Security, Ruhr-Universitaet Bochum, 44780 Bochum, Germany (email: anita.aghaie@rub.de).}\thanks{M. Mozaffari Kermani is with the Department of Computer Science and
Engineering, University of South Florida, Tampa, FL 33620, USA (e-mail:
mehran2@usf.edu).} \thanks{R. Azarderakhsh is with the Department of Computer and Electrical
Engineering and Computer Science and is an I-SENSE Fellow, Florida
Atlantic University, Boca Raton, FL, USA (e-mail: razarderakhsh@fau.edu).}}

\maketitle\pagestyle{plain}\thispagestyle{plain}
\begin{abstract}
In this paper, through considering lightweight cryptography, we present
a comparative realization of MDS matrices used in the VLSI implementations
of lightweight cryptography. We verify the MixColumn/MixNibble transformation
using MDS matrices and propose reliability approaches for thwarting
natural and malicious faults. We note that one other contribution
of this work is to consider not only linear error detecting codes
but also recomputation mechanisms as well as fault space transformation
(FST) adoption for lightweight cryptographic algorithms. Our intention
in this paper is to propose reliability and error detection mechanisms
(through linear codes, recomputations, and FST adopted for lightweight
cryptography) to consider the error detection schemes in designing
\textit{beforehand} taking into account such algorithmic security.
We also posit that the MDS matrices applied in the MixColumn (or MixNibble)
transformation of ciphers to protect ciphers against linear and differential
attacks should be incorporated in the cipher design in order to reduce
the overhead of the applied error detection schemes. Finally, we present
a comparative implementation framework on ASIC to benchmark the VLSI
hardware implementation presented in this paper.
\end{abstract}

%
%


\section{Introduction}

Research on error detection of primitives in the hardware VLSI structures
of cryptographic algorithms has been center of attention in prior
work {[}1{]}-{[}11{]}. In addition, cipher designers construct the
MixColumn transformation by a linear diffusion layer with maximum
branch number, known as maximal distance separable (MDS) matrices.
We also mention that the MDS matrices applied in the MixColumn (or MixNibble)
transformation of ciphers to protect ciphers against linear and differential
attacks should be incorporated in the cipher design in order to reduce
the overhead of the applied error detection schemes.

To motivate the urgency of embedding error detection as part of the
design cycle, we briefly go over the complications of adopting fault
FST method for lightweight ciphers (it is a motivation to our proposed
criteria in the following sections in terms of utilizing MDS matrices
as the mapping function). This method which suggests a generic FST
mapping for data storage during encryption/decryption operations of
AES-like ciphers increases the security of expensive redundancy countermeasures
against fault attacks \cite{key-1ggg}. FST is utilized to make ``fault
collision'' difficult during attacks on the classic redundancy-based
countermeasures, i.e., the adversary would have challenge in injecting
same fault in the storage registers having both original and spatial/time
redundancy structures. The
countermeasures based on recomputations could fail to detect the occurrence
of a fault as long as the adversary could inject the same fault in
both the original and redundant computations (biased fault model makes
it easier). The countermeasures based on recomputations can be used
in conjunction with encoding schemes which nullify the effect of the
bias in the fault model by FST, thwarting both these attack schemes.

To investigate the importance of using/reusing MDS matrices in lightweight
block ciphers, we have applied this method to the KLEIN cipher. Implementing
the ``naive'' spatial redundancy of KLEIN, we need 232 occupied
slices on Virtex-7 (xc7vx330t) with high area overhead. Moreover,
we have implemented the spatial redundancy with FST method that applies
MixNibble as a mapping function $W$ through this redundancy and using
InvMixNibble as its inverse $W^{-1}$. We note that although KLEIN
allows two types of decryption through (a) using encryption transformations
but utilizing modes of operations and (b) reverse transformations,
this might not be the case for other lightweight block ciphers and
that adds to the complications of using FST in lightweight cryptography.
Our implementations show that higher area, i.e., 239 occupied slices,
is achieved, as expected, due to the $W$ function. Applying the pipeline
method, which utilizes MixNibble operation as $W$ function, improves
the spatial redundancy algorithm metrics with occupying just 235 slices
on Virtex-7.

Our intention in this paper is to propose reliability and error detection
mechanisms (through linear codes, recomputations, and FST adopted
for lightweight cryptography) to consider the error detection schemes
in designing beforehand taking into account such algorithmic security. The MDS matrices applied in the MixColumn (or MixNibble)
transformation of ciphers to protect ciphers against linear and differential
attacks should be incorporated in the cipher design in order to reduce
the overhead of the applied error detection schemes.

\section{Error Detection of VLSI Architectures for MixColumn}

MixColumn has a significant role to perform as the linear diffusion
layer in the encryption and decryption operations over the finite
fields. Although, there is a wide range of categories such as circulant,
Hadamard, Cauchy, and Hadamard-Cauchy for the MDS matrices to apply
in MixColumn, choosing an efficient MDS matrix should be carefully
considered in terms of low-cost hardware area, high diffusion speed,
and low-latency implementation. One of the common methods to construct
lightweight MDS matrices, e.g., circulant, is sparing and compacting
in implementation, and then composing it several times in which it
provides similar rows in matrices to reduce the hardware implementation
cost (number of XOR gates, for instance) like Photon hash functions
\cite{key-12}.

The design criteria of MDS matrices, e.g., based on a low Hamming
weight polynomial, generate a wide pool of involutory and non-involutory
MDS matrices. Moreover, the security of these MDS matrices should
be considered carefully during the design phase to improve the security
levels.\begin{figure}
\centering\includegraphics[scale=0.35]{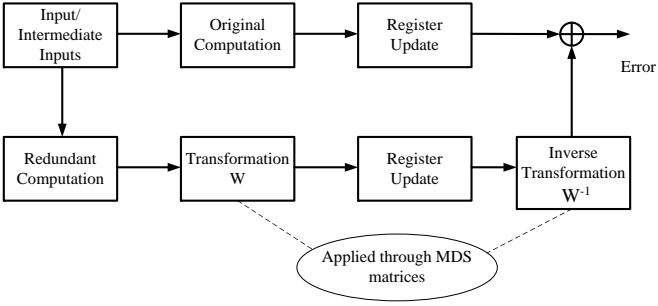}

\caption{The FST approach with MDS matrices mapping function.}
\end{figure}

All types of MDS matrices offer optimal linear diffusion to provide
the proper linear part, the MixColumn operation in block ciphers and
hash functions, but in general, a compact description for this matrix
on which one is better may not be very achievable. The criteria in
\cite{key-13i} potentially lead to low number of gates in hardware
implementations and small amount of memory usage. The $m\times m$
MDS matrix design criteria are presented in Algorithm 1.

\begin{algorithm}
\caption{$m\times m$ MDS matrix design criteria.}

{\small{}Inputs: $A:F_{2}^{m}\rightarrow F_{2}^{m}$; $X=(x_{1},...,x_{n})\in(F_{2}^{m})^{n}$, }{\small \par}

{\small{}Outputs: Criteria of MDS matrix.}\\
{\small \par}

{\small{}1. Define a linear diffusion $L(X)=(\stackrel[i=1]{n}{\sum}L_{1,i}(x_{i}),...,\stackrel[i=1]{n}{\sum}L_{n,i}(x_{i}))$,
for $1\leq i,j\leq n$, and if $L\circ L=X,$ then it is involutory.\vspace{10pt}}\\
{\small{}2. Define the bundle weight of X, $\omega_{b}(x)=|\{x_{i}:x_{i}\neq0,1\leq i\leq n\}|$,\vspace{10pt}}{\small \par}

{\small{}3. Define the branch number of L: $N=min\{\omega_{b}(X)+\omega_{b}(L(x))|X\in(F_{2}^{m})^{n},X\neq0\}$,
and if $N=n+1,$ then it is an MDS matrix.}{\small \par}
\end{algorithm}

For each of the MDS matrices, we present the multiplication and reduction
operations with irreducible polynomials to count the number of XOR
gates. The number of XOR gates for a number of lightweight block ciphers
is presented in Table I which also shows the overhead percentages
(this table is presented at the end of this section).

First cipher, Midori64, can utilize three $4\times4$ MDS matrices
for the MixColumn transformation. We investigate two of them, i.e.,
the non-involutive MDS matrix ($M_{B}$) and the involutive almost
MDS matrix ($M_{C}$). The former one is the same as the MDS matrix
applied in KLEIN (will be shown in more details), and the latter form,
$M_{C}$ , is shown below. Let us denote the input state of MixColumn
as $A$ and the output state as $R$. Then, we have the following:

$R=M_{C}\times A\Longrightarrow\begin{pmatrix}r_{0} & r_{1} & r_{2} & r_{3}\\
r_{4} & r_{5} & r_{6} & r_{7}\\
r_{8} & r_{9} & r_{10} & r_{11}\\
r_{12} & r_{13} & r_{14} & r_{15}
\end{pmatrix}=$

\begin{equation}
\begin{pmatrix}0 & 1 & 1 & 1\\
1 & 0 & 1 & 1\\
1 & 1 & 0 & 1\\
1 & 1 & 1 & 0
\end{pmatrix}\times\begin{pmatrix}a_{0} & a_{1} & a_{2} & a_{3}\\
a_{4} & a_{5} & a_{6} & a_{7}\\
a_{8} & a_{9} & a_{10} & a_{11}\\
a_{12} & a_{13} & a_{14} & a_{15}
\end{pmatrix}.
\end{equation}

According to above, we have modulo-2 added three input elements of
each column to generate each element of the output matrix ($R$),
in which each output column needs eight XOR gates. Because of the
fact that the coefficients of the input state matrix are 0 or 1, we
have the number of XOR gates as eight and twelve XOR gates for cumulative
column signature and interleaved cumulative column signature, respectively,
as shown below.

\begin{center}
$r_{0}=a_{4}+a_{8}+a_{12},$
\par\end{center}

\begin{center}
$r_{4}=a_{0}+a_{8}+a_{12},$
\par\end{center}

\begin{center}
$r_{8}=a_{0}+a_{4}+a_{12},$
\par\end{center}

\vspace{-5mm}

\begin{equation}
r_{12}=a_{0}+a_{4}+a_{8}.
\end{equation}

\noindent Let us modulo-2 add the first column of the state output
matrix to derive the cumulative column signature-based scheme.\vspace{5pt}

\noindent $r_{0}+r{}_{4}+r{}_{8}+r{}_{12}=(0+1+1+1)a_{0}+(1+0+1+1)a_{4}$\vspace{-4pt}

\begin{equation}
+(1+1+0+1)a_{8}+(1+1+1+0)a_{12}.
\end{equation}

For interleaved cumulative column signature, let us modulo-2 add two
even-row elements of the output state, i.e., rows 0 and 2, and two
odd-row elements, i.e., rows 1 and 3:\vspace{5pt}

$r_{0}+r{}_{8}=(0+1)a_{0}+(1+1)a_{4}+(1+0)a_{8}$\vspace{-4pt}

\begin{equation}
+(1+1)a_{12}=a_{0}+a_{8},
\end{equation}\vspace{5pt}

$r_{4}+r{}_{12}=(1+1)a_{0}+(0+1)a_{4}+(1+1)a_{8}$\vspace{-4pt}

\begin{equation}
+(1+0)a_{3}=a_{4}+a_{12}.
\end{equation}

As shown in Table I, we need $4\times8$ XOR gates in Midori64 with
$M_{C}$, in which the total number of cumulative column signature
gates is $4\times3$, and the required XOR gates for interleaved cumulative
column signature is $4\times2$. Due to the fact that these XOR gates
are used in all of the MixColumn transformations similarly, we do
not count them in the table.

The last cipher to present the details for the sake of brevity, LED,
applies a hardware-friendly MDS matrix for the MixColumn transformation,
that is given by:

\begin{equation}
M=\begin{bmatrix}4 & 1 & 2 & 2\\
8 & 6 & 5 & 6\\
B & E & A & 9\\
2 & 2 & F & B
\end{bmatrix}.
\end{equation}

Each entity in the input and the output state matrices is a four-bit
nibble. As a case study, to compute $r_{0}$ (the first element of
the resultant matrix $R$), denoting the bits of the elements of $A$
as $a_{ij}$ for $i$th bit of $j$th element, we have:

\begin{equation}
r_{0}=4.a_{0}+a_{4}+2.a_{8}+2.a_{12}=x^{2}.a_{0}+a_{4}+x.a_{8}+x.a_{12},
\end{equation}
\vspace{-20pt}

\[
r_{0}=x^{2}.[a_{10}x^{3}+a_{20}x^{2}+a_{30}x+a_{40}]+[a_{14}x^{3}+a_{24}x^{2}+a_{34}x
\]
\vspace{-15pt}

\[
+a_{44}]+x.[a_{18}x^{3}+a_{28}x^{2}+a_{38}x+a_{48}]+x.[a_{1c}x^{3}
\]
\vspace{-10pt}
\begin{equation}
+a_{2c}x^{2}+a_{3c}x+a_{4c}].
\end{equation}

Using the irreducible polynomial $X^{4}+X+1$ utilized for reductions,
one can derive the following for $r_{0}$:

\[
r_{0}=x^{3}.[a_{2c}+a_{28}+a_{14}+a_{30}]+x^{2}.[a_{10}+a_{40}+a_{24}+a_{38}
\]
\vspace{-15pt}

\[
+a_{3c}]+x.[a_{10}+a_{20}+a_{34}+a_{18}+a_{48}
\]
\vspace{-15pt}

\begin{equation}
+a_{1c}+a_{4c}]+1.[a_{20}+a_{44}+a_{1c}+a_{18}].
\end{equation}

Let us derive the formulae for just one column signature of MixColumn
by modulo-2 adding the first column entries $r_{0},r_{4},r_{8}$,
and $r_{12}$. One can derive:

\[
r_{4}=8.a_{0}+6.a_{4}+5.a_{8}+6.a_{12}=x^{3}.a_{0}+
\]
\vspace{-15pt}

\begin{equation}
(x^{2}+x).a_{4}+(x^{2}+1).a_{8}+(x^{2}+x).a_{12},
\end{equation}

\[
r_{8}=B.a_{0}+E.a_{4}+A.a_{8}+9.a_{12}=(x^{3}+x+1).a_{0}
\]
\vspace{-15pt}

\begin{equation}
+(x^{3}+x^{2}+x).a_{4}+(x^{3}+x).a_{8}+(x^{3}+1).a_{12},
\end{equation}

\[
r_{12}=2.a_{0}+2.a_{4}+F.a_{8}+B.a_{12}=
\]
\vspace{-25pt}

\begin{equation}
x.a_{0}+x.a_{4}+(x^{3}+x^{2}+x+1).a_{8}+(x^{3}+x+1).a_{12}.
\end{equation}

For the cumulative column signature-based scheme, we modulo-2 add
the first column entries of matrix $R$ to derive the following signature
$\hat{P}$:

\[
r_{0}+r_{4}+r_{8}+r_{12}=(4+8+B+2).a_{0}+(1+6+E+2).a_{4}
\]
\vspace{-15pt}

\[
+(2+5+A+F).a_{8}+(2+6+9+B).a_{12}
\]
\vspace{-15pt}

\begin{equation}
=5.a_{0}+B.a_{4}+2.a_{8}+6.a_{12}.
\end{equation}

After the reduction, one can derive the below formulae as the final
form:

$x^{3}[a_{10}+a_{28}+a_{44}+a_{24}+a_{3c}+a_{2c}+a_{30}]+x^{2}.[a_{10}+$\vspace{4pt}

$a_{40}+a_{34}+a_{38}+a_{20}+a_{4c}+a_{3c}+a_{1c}+a_{14}]+x.[a_{10}+a_{20}+$\vspace{4pt}

$a_{30}+a_{44}+a_{48}+a_{14}+a_{18}+a_{2c}+a_{24}+a_{4c}]+$\vspace{4pt}

\vspace{-10pt}
\begin{equation}
1.[a_{20}+a_{44}+a_{1c}+a_{18}+a_{40}+a_{34}+a_{14}+a_{2c}].
\end{equation}

This can be generalized to other columns and thus we have the followings
for the second to the fourth columns after the reductions:

\begin{equation}
r_{1}+r_{5}+r_{9}+r_{13}=5.a_{1}+B.a_{5}+2.a_{9}+6.a_{13},
\end{equation}
\vspace{-20pt}

\begin{equation}
r_{2}+r_{6}+r_{10}+r_{14}=5.a_{2}+B.a_{6}+2.a_{10}+6.a_{14},
\end{equation}
\vspace{-20pt}

\begin{equation}
r_{3}+r_{7}+r_{11}+r_{15}=5.a_{3}+B.a_{7}+2.a_{11}+6.a_{15}.
\end{equation}

In the following, the other signature-based scheme (interleaved cumulative
column signature) is derived through modulo-2 adding the odd-row elements
with each other and the even ones as well:

\begin{equation}
r_{0}+r_{8}=F.a_{0}+F.a_{4}+8.a_{8}+B.a_{12},
\end{equation}

\begin{equation}
r_{4}+r_{12}=A.a_{0}+4.a_{4}+A.a_{8}+D.a_{12}.
\end{equation}

According to these formulae, we are able to count the number of utilized
XOR gates and the cumulative column signature and interleaved cumulative
column signature overheads for LED which are presented in Table I.
As mentioned above, by default, for each cipher, we need $12$ and
$8$ XOR gates for cumulative column signature and interleaved cumulative
column signature, respectively, in addition to what presented, which
are omitted for the sake of brevity.
\begin{table}
\caption{Number of gates needed for the MixColumn transformation and deriving
the predicted signatures in different lightweight block ciphers}

\centering%
\begin{tabular}{|c|c||c||c|}
\hline
\multirow{2}{*}{Block cipher} & \multicolumn{1}{c||}{MixCol.} & \multicolumn{1}{c||}{CCS} & \multicolumn{1}{c|}{Inter. CCS}\tabularnewline
\cline{2-4}
 & XOR & XOR & XOR\tabularnewline
\hline
\hline
\multicolumn{1}{|c|}{Midori64 ($M_{C}$)} & 128 & 176 (37.50\%) & 160 (25\%)\tabularnewline
\hline
Midori64 ($M_{B}$) & 256 & 304 (18.75\%) & 416 (62.50\%)\tabularnewline
\hline
LED & 444 & 564 (27.02\%) & 672 (51.35\%)\tabularnewline
\hline
KLEIN (two-nibble) & 256 & 304 (18.75\%) & 416 (62.50\%)\tabularnewline
\hline
\end{tabular}
\end{table}

Finally, as mentioned before, we summarize the number of XOR gates
for all the mentioned ciphers in the MixColumn transformation in Table
I, in which the overhead percentages are presented. We have not used
sub-expression sharing, similar to the S-boxes, for deriving the numbers
in Table I; nonetheless, sub-expression sharing can be used to reduce
the number of gates at the expense of possible high fan-outs (which
might not be tolerated in some cases, requiring repeaters to resolve
the problem). Table I and the MixColumn transformations that we have
not considered for the sake of brevity motivate the urgency of considering
the overheads beforehand, perhaps as a design factor (we note that
other error detection schemes can be considered and the ones provided
here are just a subset).

Similar to the S-boxes, we present two metrics for analyzing the results
in Table I. The first one is the overhead for our error detection
schemes shown in Table I. As seen in this table, Midori64, which applies
the $M_{C}$ matrix, has the lowest-cost for the MixColumn implementation;
nevertheless, the cumulative column signature-based scheme overhead
of this matrix is more than other ones. Comparing the percent overheads
in Table I shows how different they could be with respect to error
detection using cumulative column signature (and other schemes for
error detection). The second metric is the total number of gates (the
original and the add-on detection), e.g., the number of applied logic
gates of $M_{B}$ is not as low as $M_{C}$.
\begin{table}
\caption{ASIC TSMC 65-nm synthesis results for two select MixColumn transformations
and their error detection mechanisms (cumulative column signatures:
CCS)}
\centering{}{
\begin{tabular}{|c||c|c|c|}
\hline
\multirow{2}{*}{Block Cipher} & \multicolumn{3}{c|}{GE of Architectures}\tabularnewline
\cline{2-4}
 & MixCol. & CCS & Inter. CCS\tabularnewline
\hline
\hline
\multirow{1}{*}{Midori, $M_{C}$ } & 272 & 330 (21.3\%) & 317 (16.5\%)\tabularnewline
\hline
LED & 575 & 746 (29.7\%) & 765 (33.0\%)\tabularnewline
\hline
\end{tabular}{}}{ \par}
\end{table}

Finally, through ASIC synthesis and for two select constructions in
Table II, we present the areas for the MixColumn transformations of
Midori ($M_{C}$) and LED. The benchmarking is done for the error
detection architectures using TSMC 65-nm library and Synopsys Design
Compiler. Similar to the S-boxes, in order to make the area results
meaningful when switching technologies on ASIC, we have provided the
NAND-gate equivalency (gate equivalents: GE). This is performed using
the area of a NAND gate in the utilized TSMC 65-nm CMOS library which
is 1.41 $\mu m^{2}$. The results are shown in Table II, where the
overheads are presented in parentheses (the contrast when comparing
this table and Table I is because of the optimizations performed in
Design Compiler, noting that we have not performed sub-expression
sharing in Table I). The aforementioned metrics/overheads are some
of the possible indications to give designers the required criteria
to predict low-cost MixColumn implementations for lightweight ciphers.

\section{Conclusions}

In this paper, we evaluated the hardware complexities of the MixColumn
transformations to propose a framework for design-for lightweight
and effective countermeasures for intentional and natural faults in
crypto-architectures and to present respective motivations. One can
also base the criteria (depending on the objectives) on other performance
and implementation metrics, e.g., delay (frequency, throughput, efficiency),
power consumption, and energy. Although we chose the MixColumn transformation
due to their importance, other less costly transformations can be
considered. The results of our VLSI implementations on ASIC platform
shows the diversity of MixColumn in lightweight cryptography, calling
for efficient approaches for error detection. Finally, one could consider
a subset of fault attacks, differential fault intensity analysis (DFIA),
see for instance, \cite{key-3,key-5-1,key-6-1}, which combines differential
power analysis with fault injection principles to obtain biased fault
models (multi-byte faults cannot be used practically for attacking
time redundancy countermeasure implementations, and single-byte fault
models are the only viable option for the attackers).

\end{document}